\documentclass[openacc]{rsproca_new}

\usepackage{arxiv}

\usepackage[utf8]{inputenc} 
\usepackage[T1]{fontenc}    
\usepackage{hyperref}       
\usepackage{url}            
\usepackage{booktabs}       
\usepackage{amsfonts}       
\usepackage{nicefrac}       
\usepackage{microtype}      
\usepackage{graphicx}
\usepackage{doi}
\usepackage[numbers]{natbib}
\usepackage{xcolor}
\usepackage{comment}
\usepackage{tabularx}

\title{Scientific Applications of Quantum Computing: Challenges and Opportunities}

\author{
\normalfont
Bruno Camino$^{[1]*}$, C.~Richard~A.~Catlow$^{[1,2,3]*}$, John Buckeridge$^{[4]}$, Alin~M.~Elena$^{[5]}$,\\ 
Vladimir V. Gusev$^{[6]}$, Sarah Harris$^{[7]}$, Thomas~W.~Keal$^{[5]}$, Glenn Jones$^{[8]}$, Vivien Kendon$^{[9]}$, Syma Khalid$^{[10]}$,\\ Phalgun Lolur$^{[11,12]}$,
Jamal~A.~Nasir$^{[1]}$, Matthew~J.~Rosseinsky$^{[13,14]}$, Chris-Kriton Skylaris$^{[15]}$,\\
Paul~A.~Warburton$^{[16]}$ and Scott~M.~Woodley$^{[1]*}$ \\
\\[0.75em]
$^{1}$Department of Chemistry, Kathleen Lonsdale Building, University College London,\\
20 Gordon Street, London WC1H 0AJ, United Kingdom \\
$^{2}$Cardiff Catalysis Institute, School of Chemistry, Cardiff University,\\ Park Place, Cardiff CF10 3AT, United Kingdom \\
$^{3}$UK Catalysis Hub, Research Complex at Harwell,\\ Rutherford Appleton Laboratory, Harwell OX11 0QX, United Kingdom \\
$^{4}$School of Engineering and Design, London South Bank University,\\ 103 Borough Rd, London SE1 0AA, United Kingdom \\
$^{5}$Scientific Computing Department, STFC Daresbury Laboratory, Warrington,\\ Cheshire WA4 4AD, United Kingdom \\
$^{6}$School of Computer Science and Informatics, University of Liverpool,\\ Liverpool L69 3DR , United Kingdom\\
$^{7}$School of Mathematical and Physical Sciences, University of Sheffield,\\ Hounsfield Road, Sheffield S3 7RH, United Kingdom \\
$^{8}$Phasecraft Ltd., London W1T 4PW, United Kingdom \\
$^{9}$Department of Physics, University of Strathclyde, Glasgow G4 0NG, United Kingdom \\
$^{10}$Department of Biochemistry, University of Oxford, Oxford OX1 3QU, United Kingdom\\
$^{11}$Capgemini UK PLC, Engineering Science, 95 Queen Victoria Street, London EC4V 4HN, United Kingdom \\
$^{12}$Capgemini Quantum Lab, Place de l'\'Etoile, 11 rue de Tilsitt, 75017 Paris, France \\
$^{13}$Department of Chemistry, University of Liverpool, Crown Street, Liverpool L69 7ZD, United Kingdom \\
$^{14}$Leverhulme Research Centre for Functional Materials Design, Materials Innovation Factory,\\ University of Liverpool, 51 Oxford Street, Liverpool L7 3NY, United Kingdom \\
$^{15}$School of Chemistry and Chemical Engineering, University of Southampton,\\ Highfield, Southampton SO17 1BJ, United Kingdom \\
$^{16}$London Centre for Nanotechnology, University College London,\\ London WC1H 0AH, United Kingdom \\
\\[0.5em]
\small
$^{*}$Corresponding authors: b.camino@ucl.ac.uk; c.r.a.catlow@ucl.ac.uk; scott.woodley@ucl.ac.uk
}
\date{}

\renewcommand{\shorttitle}{\textit{arXiv} Template}

\hypersetup{
pdftitle={A template for the arxiv style},
pdfsubject={},
pdfauthor={},
pdfkeywords={Quantum Computing, Chemistry, Materials Science, Biology},
}

\begin{document}
\maketitle
\newpage

\begin{abstract}
	The predictive simulation of molecules and materials has had a broad and significant impact. It nevertheless remains constrained by the cost of accurately treating electronic correlation, excited states, and complex energy landscapes. Quantum computing offers a fundamentally different computational paradigm in which quantum states are encoded and manipulated directly rather than approximated on classical hardware. Here we discuss where this approach may provide a genuine scientific advantage in chemistry, materials science, and biochemistry. Promising directions include the high-accuracy treatment of correlated active spaces, improved excited-state simulations, and accelerated exploration of combinatorial structure spaces. The central challenge is therefore not qubit scaling alone, but demonstrably chemically meaningful gains in predictive reliability. We argue that near-term value is most likely to come from disciplined workflow integration rather than wholesale replacement of classical methods. Noisy physical devices, error-mitigated utility experiments, early fault-tolerant devices, and fully fault-tolerant quantum computers offer different scientific prospects, and claims of usefulness must be tied to the specific regime being discussed. Quantum computing will become scientifically valuable when it demonstrably reduces uncertainty in computed energies, rates, spectra, or materials stability after the full costs of state preparation, measurement, error handling, and coupling to classical simulation are included.
\end{abstract}


\section{Introduction}
Chemistry and materials science are integral to the development of civilisation \cite{dobrzanski2006significance}. From pigments and metallurgy to ceramics, catalysts, energy materials, and electronics, progress has followed a pattern of empirical discovery, theoretical description, and eventually predictive modelling \cite{kumar2025multidisciplinary}. Over the past few decades, computation has become an essential part of this process. By encoding physical theory into algorithms, computer simulations allow chemical behaviour to be analysed and, in some cases, predicted \cite{krzywanski2024advanced}. \\

The impact of computational modelling has been driven by three parallel advances. The first is hardware: the exponential growth of high-performance computing has enabled simulations of increasing size and accuracy \cite{vetsch2025ab}; the second is fundamental theory, which has allowed increasingly accurate simulations of matter at the atomic and molecular level; the third is methodological: algorithmic developments have reduced computational costs and extended the range of problems that can be addressed~\cite{garcia2012review}. Together, these advances have made computer simulations a routine component of research across chemistry, materials science, and molecular biology \cite{shahzad2024accelerating,baiardi2023quantum}. Despite this progress, fundamental limitations remain: many problems of practical interest involve electronic correlation, excited states, or rare events that are difficult to describe reliably with existing methods \cite{vetsch2025ab,Stocks2024}. \\

Quantum computing represents a paradigm shift in computational capabilities that relies on quantum mechanical phenomena such as superposition, interference, and entanglement \cite{AuYeung2024}. This emerging technology holds promise for solving complex problems in materials science, chemistry, and biochemistry that are difficult for current classical approaches in selected regimes \cite{weidman2024quantum,paudel2022quantum,marchetti2022quantum}. The potential impact of quantum computing across various sectors, for example, healthcare and life sciences, is significant \cite{flother2023state}. Importantly, quantum computing should not be viewed in isolation from existing computational approaches. Classical methods continue to improve, from tensor-network electronic-structure techniques and machine-learned interatomic potentials to exascale electronic-structure/molecular-dynamics software and enhanced-sampling methods for rare events~\cite{wang2024multi, ansari2026convergence, di2023perspective, he2025machine}. Since chemically relevant systems typically span multiple length and time scales, quantum computing is likely to be most useful as part of a hybrid computational ecosystem rather than as a standalone replacement. The relevant question is therefore not whether quantum computers will replace classical ones, but how they might be integrated into established workflows to address specific computational bottlenecks. We expect quantum computers to act as highly specialised accelerators, working hand-in-hand with classical methodologies.\\

Quantum computing algorithms can be divided into two overarching paradigms dependent upon whether the quantum states evolve in continuous time or discrete time. Both approaches may be used for the simulation of quantum materials and quantum chemistry. In the case of simulation using a continuous-time quantum computer, the Hamiltonian that describes the dynamics of the system to be modelled is implemented directly (or at least approximately) on the quantum computer. This paradigm has much in common with quantum annealing, in which computational problems are encoded into a Hamiltonian whose ground state is found by continuous evolution. Such continuous-time approaches are often referred to as analogue quantum computation. On the other hand, discrete-time quantum simulation (or equivalently, digital quantum simulation) makes use of quantum gates. Here, the connection between the Hamiltonian of the system being simulated and the sequence of gates that perform the simulation is not direct. The gate set may need to be found iteratively using hybrid quantum/classical approaches such as the variational quantum eigensolver (VQE) \cite{WOS:000867093900001} and the Quantum Approximate Optimization Algorithm (QAOA) \cite{WOS:001221953700001}, also generalised as the quantum alternating operator ansatz \cite{hadfield2019fromqaoa}. Separately, Trotter--Suzuki product formulas provide a general method for approximating the time evolution of a Hamiltonian as a sequence of gates, and underpin most gate-based approaches to digital quantum simulation~\cite{lloyd1996universal}. A related but distinct approach, digitised adiabatic simulation, implements an annealing schedule as a gate sequence rather than running it continuously. Fault-tolerant gate-based algorithms, such as quantum phase estimation, represent a further distinct regime in which quantum error correction is required to support circuit depths beyond those achievable on present noisy hardware.\\

Much like classical computation, discrete-time quantum computation benefits from (and indeed is largely dependent upon) the existence of several proposals for error correction. Error correction is also relevant to continuous-time and analogue approaches, particularly when the goal is long, complex, programmable computations with reliable outputs, although comparatively few such schemes have so far been proposed \cite{WOS:000332666700001}. For specific problems, present-day analogue and annealing-based simulators have outperformed present-day gate-based devices: programmable neutral-atom arrays have realised many-body dynamics and phase transitions with several hundred qubits \cite{WOS:000671377900009,WOS:000671377900010,WOS:000728578700055}, and quantum annealers have reached several thousand qubits \cite{WOS:000975071100001}, exceeding the scale of comparable gate-based demonstrations. Gate-based approaches, however, are advancing rapidly on a different axis: resource estimates for fault-tolerant algorithms have fallen substantially and logical qubits with reduced error rates have now been demonstrated experimentally \cite{googlequantumai2025colourcode,gidney2024magicstate,gidney2025factor}. It remains to be seen how these trajectories compare as the technology matures, and whether gate-based quantum simulation (and computation) will eventually consign present-day analogue approaches to history, as has largely happened with simulation by classical analogue electronics.\\

In this perspective, we examine where quantum computing may contribute to molecular and materials modelling; also where it is unlikely to help and what conditions must be met for it to become practically useful. Rather than surveying all proposed algorithms or platforms, we focus on the challenges posed by realistic chemical systems and assess how quantum approaches might fit within the broader landscape of scientific computing.\\

\section{Current challenges in molecular and materials modelling}
We briefly outline the dominant sources of uncertainty across chemistry, materials science, and biochemistry, focusing on where computational cost and accuracy remain limiting in practice.

\subsection{Chemistry}
The central open question for computational chemistry is whether electronic-structure methods can deliver chemical accuracy for realistic, multi-component, dynamic systems, rather than only for small or idealised models, without prohibitive computational cost. Quantum resources could realistically contribute where classical methods are most strained: high-accuracy treatment of strongly correlated active spaces, excited-state and spectroscopic properties, and benchmark energies that feed back into cheaper classical methods.\\

At the heart of chemistry and materials science lies a deceptively simple objective: predicting thermodynamics and kinetics with sufficient accuracy to explain and ultimately control chemical behaviour \cite{le2012quantitative}. Whether one considers enzymatic catalysis, solid-state decomposition, heterogeneous catalysis, or molecular recognition, macroscopic observables are governed by free-energy differences and activation barriers that originate at the electronic scale \cite{blumberger2015recent} \cite{haber1997molecular}. The central challenge is not conceptual but quantitative: these observables depend exponentially on energy differences, so even modest errors at the atomistic level can propagate into orders-of-magnitude uncertainty in predicted rates or equilibria. This sensitivity places stringent demands on electronic-structure methods. Achieving chemical accuracy, roughly a few tens of meV in reaction or activation free energies, remains difficult for anything beyond the smallest systems \cite{lewis2022machine}. Density functional theory (DFT)~\cite{jones2015density}, the workhorse of modern materials modelling, offers an appealing balance between cost and accuracy but lacks systematic error control; fortuitous cancellation of errors may underpin some of its success. Highly correlated wavefunction methods~\cite{nagy2024state} can, in principle, deliver the required precision, yet their steep computational scaling places them beyond reach for chemically realistic systems, even on emerging exascale classical hardware. As a result, although there are truly predictive simulations of reaction energetics and kinetics, for example the work of Piccini \textit{et al.} on zeolite catalysed reaction \cite{piccini2016ab}, these remain the exception rather than the rule.\\

 Quantum computing enters this landscape not as a complete replacement for classical approaches, but as a potential complement \cite{kitaev2002classical}, thereby allowing us to go beyond high-level correlated quantum chemical methods and one day outperform the beyond-DFT methods
 \cite{rayabharam2025hybrid} \cite{ho2018promise}. Quantum computing is not expected to supplant classical methodologies in materials modelling. Instead, it will function as a complementary tool, applied where it offers a clear advantage while remaining embedded within broader multi-scale modelling strategies. The intrinsic complexity of real materials ensures that progress will continue to depend on integrated, multi-method approaches that couple quantum mechanical accuracy with hierarchical modelling frameworks \cite{sun2013review}. Furthermore, even with perfect electronic-structure solvers, chemistry does not reduce to isolated molecules at zero temperature. Real materials are complex, heterogeneous, and dynamic. Industrially relevant systems are multi-component, often far from equilibrium, embedded in solvents or electrolytes~\cite{Dziedzic2020}, exposed to applied fields, photons, or reactive environments, and evolve over time scales ranging from femtoseconds to years \cite{goddard2020multi,hu2023ultrafast}. It is unreasonable to expect any single method to solve all of these problems. We will see quantum devices solving problems that quantum computers are good for, and classical devices solving problems best suited to their architecture. \\

Chemical accuracy is required not only for energies but also for free energies, demanding extensive exploration of high-dimensional phase spaces \cite{donoho2009observed}. Molecular dynamics, Monte Carlo, and enhanced-sampling techniques dramatically increase computational costs and are often intrinsically serial, limiting scalability \cite{lazim2020advances,bernardi2015enhanced}. While alternative, more parallel sampling strategies are emerging, efficient and reliable exploration of complex reaction networks, especially in biochemical or catalytic environments, remains an open challenge. Composition and structural complexity further compound the problem. Predicting behaviour across vast compositional spaces requires both reliable interaction models and the ability to explore configurational landscapes involving disorder, segregation, and phase competition \cite{batten2001complex}. Classical approaches such as cluster expansions and Monte Carlo methods have proven powerful, but their accuracy ultimately rests on underlying electronic-structure data \cite{cao2018use}. Here again, quantum computing could play a role by supplying benchmark energies or by enabling new algorithmic approaches to landscape exploration. Spectroscopy provides a complementary perspective on these challenges. Many experimental probes interrogate excited states, electron–phonon coupling, or light matter interaction regimes that stretch classical simulation methods beyond their natural domain \cite{kilina2015light}. While approximate classical treatments remain essential, quantum devices offer, in principle, a more direct route to excited-state and spectroscopic simulations by implementing quantum phase estimation for excitation energies, quantum subspace/Krylov methods for low-lying excited states, and real-time Hamiltonian-simulation or correlation-function algorithms for absorption, emission, vibronic, and nonlinear spectroscopies~\cite{aspuru2005simulated, wang2008quantum, mcclean2017hybrid, ollitrault2020quantum}. Such methods could be particularly useful for photocatalysis, energy conversion, and optoelectronic materials, where excited-state lifetimes, charge-transfer processes, and light–matter response are key observables.\\

Finally, these scientific challenges intersect with practical constraints. Computational chemistry generates vast data sets whose storage, curation, and analysis are increasingly non-trivial, while high performance computing remains energy intensive. Developing scalable, energy-efficient algorithms and architectures (be they classical, quantum, or hybrid) will be essential if predictive chemical simulation is to become routine rather than aspirational \cite{abas2024high}. To this end, the grand challenge is not simply to compute more accurately, but to integrate methods across scales, architectures, and disciplines. Quantum computing may ultimately reshape parts of this landscape, but its success will depend on how effectively it is woven into the broader fabric of computational chemistry rather than on isolated demonstrations of advantage \cite{panda2024quantum}. Quantum computing offers the prospect of more faithful representations of correlated electronic structure and quantum dynamics, potentially addressing some of the fundamental limitations of classical methods~\cite{McArdle2020}. Assessing how such approaches may be integrated into realistic simulation workflows, and the timescales over which practical advantages may emerge, is a central question for the field.\\

\subsection{Materials Science}
The central open question for materials discovery is whether predictive, \textit{a priori} design can replace the intuition-driven, high-throughput experimentation that has historically delivered the field's biggest successes, given that both accurate electronic structure and tractable search over combinatorial composition and structure space are needed simultaneously. The realistic quantum contribution here is benchmark electronic-structure data for otherwise intractable systems, and accelerated exploration of the combinatorial structure and disorder spaces that classical search struggles to cover.\\

Before discussing the challenges at the interface of materials science and quantum computing, we will consider a canonical example that illustrates the intrinsic difficulty of prediction in materials science: the development of industrial ammonia synthesis~\cite{Smil2001}. 
In the early twentieth century, Haber and Bosch demonstrated nitrogen fixation using a sub-optimal osmium catalyst~\cite{Ertl2010HaberBosch}, which was followed by the extensive experimental work of P. A. Mittasch, who systematically explored tens of thousands of candidate materials and ultimately identified an iron-based catalyst, variants of which remain in industrial use~\cite{Ertl2008}. Subsequent decades have seen major advances in computational catalysis. In particular, DFT, combined with high-performance computing and experiment, has been used to rationalise the activity of Fe and Ru catalysts and to assess trends across alloy compositions~\cite{Norskov2004,Norskov2005,Medford2015}. Despite this progress, the Mittasch catalyst itself is a complex multi-component system, comprising iron oxides (Fe$_3$O$_4$) together with promoters and additives such as K$_2$O, CaO, Al$_2$O$_3$, and SiO$_2$~\cite{Ertl2008}. Even today, it is unclear whether existing computational approaches could have predicted such a catalyst a priori, including not only the active phase but also the choice and relative proportions of promoters, supports, and synthesis pathways required to generate the active material~\cite{Crawford2007}. While modern simulations can increasingly guide experimental efforts across catalysis, batteries~\cite{Morgan2022}, photovoltaics, and functional materials~\cite{Norskov2009}, predictive discovery remains limited. Recent machine-learning-based approaches have shown promise in screening large chemical spaces~\cite{Butler2018,Schmidt2019}, yet none have demonstrably exceeded the empirical success of intuition-driven, high-throughput experimentation exemplified by Mittasch’s work. \\

Within materials and chemistry simulation, key difficulties arise from both electronic-structure accuracy and multiscale complexity~\cite{Burke2012,Cao2019,Bauer2020,Babbush2019}. Many technologically relevant problems, including heterogeneous catalysis, electrochemistry, energy materials, and drug discovery, are governed by subtle differences in electronic structure, often involving strong correlation, redox processes, charge transfer, and bond breaking and formation~\cite{Cohen2012}. Small errors in computed energies can therefore lead to large uncertainties in predicted reaction rates, selectivities, or phase stability. In addition, these problems are intrinsically multiscale. Localised quantum-mechanical phenomena at active sites, defects, or transition states must be described consistently within extended environments such as bulk solids, solvents, or electrochemical interfaces~\cite{Greeley2016}. Classical computational approaches typically rely on a combination of approximations, both in the energy and structural models, and heuristic strategies to address this complexity, limiting their predictive power despite their success in rationalising trends and guiding experiments.\\

Functional materials underpin modern technologies, and the discovery of new compounds is essential for addressing global challenges such as sustainable energy production. Structure prediction plays a critical role in this process, as atomistic structure determines both stability and material properties. For crystalline materials, long-range order is typically described using a periodic unit cell, enabling the application of crystal structure prediction (CSP) methods~\cite{Crawford2019}. CSP algorithms can identify energetically viable structures and compositions, allowing competing phases to be assessed computationally~\cite{Vasylenko2021,Hunnisett2024}. However, the widespread adoption of CSP is limited by its computational cost. Exact evaluation of interatomic interactions is expensive due to their quantum-mechanical nature, while optimisation over atomic configurations within a unit cell leads to a combinatorial explosion of possibilities. Indeed, several formulations of CSP are known to be NP-complete~\cite{Adamson2022}. In practice, these challenges are addressed using classical force fields and heuristic optimisation strategies~\cite{Woodley2008,Woodley16}, which have enjoyed considerable success but may be limited in accuracy and transferability.

\subsection{Biochemistry}

The central open question for biomolecular simulation is whether multiscale models can reach the system size, timescale, and accuracy needed for quantitatively useful drug design and mechanistic insight, given the cost of simulating hundreds of thousands to millions of particles with explicit long-range electrostatics. A more realistic role for quantum resources is treating small, localised, strongly correlated regions, such as a catalytic site or a bond-breaking event, more accurately within an otherwise classical multiscale model, and accelerating specific costly classical subroutines such as electrostatics evaluation.\\

Biomolecular simulations have the potential to provide \textit{in silico} insight into the molecular causes of disease and to support the design of new medicines and therapeutic interventions. However, the computational description and representation of biomolecules are extremely challenging, despite them predominantly containing elements that are close to the top of the periodic table with a relatively small number of electrons. Biomolecules are aperiodic, so many of the simplifications used in condensed matter physics to model lattices do not apply. Individual proteins can contain many thousands of atoms, and biomolecular assemblies may contain millions~\cite{Karplus02,Shaw2010}. Biomolecules are soft matter objects that function at ambient temperatures ($\sim$300~K), so thermal fluctuations cannot be neglected and, in fact, make a vital contribution to the operation of these functional building blocks of living systems~\cite{Henzler2007}. Biological assemblies are not only large and structurally complex, but their dynamics and biological function are also strongly shaped by the surrounding molecular environment \cite{wei2016protein}. This environment includes water, salts, lipid and protein-containing membranes, metabolites, and molecular crowders~\cite{Dill2012}. Neglecting these effects can lead to unrealistic simulations, but including them often increases system sizes to hundreds of thousands, or even millions, of particles.\\

Biomolecular dynamics and processes are inherently multiscale, both temporally and spatially. Muscle contraction, for example, arises from the coordinated action of nanoscale molecular motors whose conformational cycles are powered by ATP hydrolysis. The macroscopic biological process itself is not a realistic target for direct quantum-computer simulation; rather, any potential role for quantum computing would be confined to localised electronic-structure problems, such as chemical bond rearrangement in ATP hydrolysis or enzyme-catalysed reaction \cite{babcock2025physical, liu2002probing}. Long-range electrostatic interactions are explicitly included in simulations featuring all-atom descriptions of biomolecules, which incur significant computational expense. Fast Fourier transform (FFT)-based methods, such as particle--mesh Ewald schemes, are therefore a critical computational bottleneck for current methods on classical computers and have been identified as a potential area where quantum algorithms could offer advantages in specific subroutines~\cite{Harris2010}.\\

Biomolecular simulations, therefore, represent a particularly challenging example of multiscale complexity and would benefit substantially from a step-change technological breakthrough enabling simulations that are orders of magnitude faster. Even supercomputers designed with bespoke hardware specifically to address these challenges, such as the \textit{Anton} architecture, are only now beginning to deliver sufficient computational power to make in silico drug design quantitatively useful in industry~\cite{Shaw2014}. While such approaches remain proprietary and are not yet widely accessible~\cite{DEShawDrug}, they nevertheless demonstrate the scale of computational resources required and the potential impact of further advances.

\section{HPC and QC to tackle calculations on realistic chemical systems}
Classical high performance computing (HPC) has long been established as an essential tool for the modelling and simulation of materials and molecular systems. This trend continues into the exascale computing era as increasing computing power drives ever more accurate calculations with greater fidelity to experiments and predictive power~\cite{materials_exascale_2022}. The need for large scale computing resources stems from the underlying quantum mechanical theories that describe the chemical and physical processes of interest. The scaling of these methods to model the electronic structure of systems in realistic experimental conditions (that is, with a large number of atomic centres) rapidly becomes challenging on even the largest classical computers. As a result, much of the methodological development in this area focuses on reducing computational expense while maintaining or improving accuracy.\\

Electronic structure calculations are of special interest in the development of quantum computing technology, as, in principle, quantum computers are better suited than their classical counterparts to the task of describing the inherently quantum mechanical nature of electrons and their interactions. A wide range of algorithms has been developed for solving electronic structure problems on quantum hardware, including those, such as the variational quantum eigensolver (VQE)~\cite{Peruzzo2014}, that take a hybrid approach, leveraging the processing power of classical computing to help overcome the limitations of current noisy intermediate-scale quantum computers, and which can be employed for fundamental computational chemistry tasks, such as exploring potential energy surfaces~\cite{hartree-fock_2020}. Nevertheless, severe limitations remain on the size of problems that can be tackled on current quantum computing hardware, due both to the number of qubits required as system size increases and to the depth of the quantum circuits involved.\\

As quantum computing technology is further developed, we can expect that increasingly large quantum chemistry problems will become tractable, and the full potential of this new frontier will eventually be realised for challenges across chemistry, materials science, and biomolecular modelling. However, the question remains whether we can do useful science with today’s quantum computing technology and the platforms that will follow in the short to medium term. To achieve this, we need to consider further how classical HPC can be used in combination with quantum hardware to bring simulations of complex chemical systems within reach.\\

A widely used approach for mitigating the computational cost of electronic structure calculations on classical computers is to divide the system into two parts: an inner region, typically the site where a reaction takes place, which requires a full quantum mechanical treatment to be described with sufficient accuracy, and an outer region containing the chemical environment, which can be treated using a more approximate classical method. This approach, known as quantum mechanical/molecular mechanical (QM/MM) modelling, is an example of a multiscale method as it combines calculations at the electronic and atomistic scales. It allows the accuracy of the calculation to be maintained while reducing overall computational cost compared to a purely quantum chemical calculation on the same system. QM/MM methods implemented in software such as the multiscale computational chemistry environment ChemShell~\cite{chemshell_2014,pychemshell_2019} are routinely used for modelling complex chemical systems in realistic conditions, particularly for investigating enzymatic reactivity and catalytic materials~\cite{pccp_chemshell_2023}. QM/MM embedding schemes can also be generalised to three (or more) layers~\cite{svensson_oniom_1996}. This flexibility has motivated a growing, active body of work on multiscale embedding for quantum computing, in which a calculation on a small, strongly correlated system is embedded inside a QM/MM model evaluated using classical HPC, giving a multi-layer ``QC/QM/MM'' approach capable of scaling to large system sizes. Proposals and implementations of this approach include multilayer embedding schemes aimed at pharmaceutical applications~\cite{izsak_active_space_2023}, hybrid HPC-QC frameworks for multiscale chemical modelling~\cite{Thacker2026}, bootstrap-style multiscale embedding linking QM/MM and quantum-computed fragments~\cite{weisburn2025multiscale}, quantum-classical partitioning strategies for biochemical systems~\cite{cheng2020application}, and multiscale quantum algorithms that explicitly couple quantum and classical layers~\cite{ma2023multiscale}. Such approaches are particularly well suited to current Noisy Intermediate-Scale Quantum (NISQ) devices, as they confine quantum resources to the smallest, most strongly correlated region while allowing classical HPC to handle the remainder.\\

The resulting hybrid HPC-QC approach can scale to many thousands of atoms while retaining the advantages of quantum algorithms over conventional QM methods for achieving increased accuracy in the innermost layer. It can also incorporate active space embedding methods that target a specific set of orbitals at the quantum computing level, further increasing the range of systems that can be studied~\cite{izsak_active_space_2023}. The multiscale framework offers a route to using NISQ and error-mitigated quantum computing devices for pharmaceutical and catalytic applications and, more generally, to modelling any complex chemical system~\cite{Thacker2026}. It is also highly flexible and can adapt as quantum technology progresses, with the potential for the inner QC layer to cover a larger region as the available qubits and support for increased quantum circuit depth allow.

Several issues nonetheless remain unresolved before QC/QM/MM can be considered a mature methodology: the principles for selection of the active space and of its boundary with the surrounding QM/MM region; errors introduced at that embedding boundary, including truncation and basis-set artefacts; treatment of polarisation and electrostatic response between the quantum algorithm, HPC QM, and HPC MM layers; the scheme used to couple the quantum and classical layers, and its convergence; the measurement overhead required to extract energies, gradients, or properties from the quantum layer at each optimisation or dynamics step; and resource estimation for hybrid HPC-QC calculations, including scaling of the individual component calculations.

\section{Quantum Hardware}

At present, there are several hardware platforms on which proofs-of-principle of quantum computation and quantum simulation have been performed. Each hardware platform has its respective strengths and weaknesses by comparison with the other platforms \cite{WOS:001631084900003}. More specifically, each platform has at least one critical engineering bottleneck that must be overcome if it is to become the basis of a quantum simulation or quantum computation tool at a commercially-viable scale. 

For use in the gate-based paradigm, the essential challenge for any hardware platform is to isolate each qubit sufficiently from its environment so that coherence is maintained over useful timescales, while also allowing interactions between qubits so that gates with two (or more) input qubits can be implemented. In the quantum annealing paradigm, it was originally believed that coherence played a less significant role than for gate-based quantum computation. Recent results from D-Wave, however, suggest that coherence is essential to obtaining a scaling advantage over classical benchmarks \cite{WOS:000854051400003}.

Although the implementation details differ between the gate-based and quantum annealing paradigms, several physical platforms used for quantum computing have the potential to support both approaches. At present, the leading platforms for gate-based approaches to quantum simulation are neutral atoms and superconducting transmons, while for annealers, the leading platforms are neutral atoms and superconducting flux qubits. Other hardware platforms (based on trapped ions \cite{WOS:000655978400001}, semiconductor devices \cite{WOS:001679923300001}, linear optics \cite{WOS:000599959400045, WOS:000360388400016}, or other as yet unrealised technologies) may well, in due course, outperform the current state-of-the-art.


In quantum computation and quantum simulation using neutral atoms, quantum information is typically encoded into the Rydberg states of atoms, which are held in optical tweezers (for a review, see for example \cite{WOS:000717415600001}). Coherence lifetimes are long due to the identical nature of each atomic qubit. Cycle times are also long, however. The current state of the art in terms of qubit number is of the order of thousands \cite{WOS:001603575100001}. Scaling to around 100,000 Rydberg qubits is potentially feasible; this being practically limited by how many optical tweezers (generated by a spatial light modulator) can be brought to a focus by a single optical microscope. Demonstrations of quantum simulations using neutral atoms include those of phase transitions in an Ising anti-ferromagnet \cite{WOS:000671377900009, WOS:000671377900010} and topological spin liquids \cite{WOS:000728578700055}, in both cases with over 200 Rydberg qubits. 

The leading superconducting qubit platforms are based on transmons (for the gate-based paradigm) or flux qubits (for quantum annealing). For a review of superconducting qubit technology, see for example \cite{WOS:000520427200017, WOS:000698590900001}. While gate operation times are orders of magnitude shorter for transmon qubits (by comparison with atomic or ion-trap qubits), so too are the coherence lifetimes, with the current transmon state-of-the-art being on the order of 0.1 ms \cite{saklakov2025microgravity}. Gate-based circuits containing on the order of a few hundred qubits and annealing circuits containing a few thousand qubits have been demonstrated in both computational and simulation contexts. Scaling to larger systems is presently limited by the need for multiple wires from room-temperature to the quantum processor at 10 mK. Such wiring acts as both a heat load and a source of decohering noise. Instances of using such superconducting processors to perform quantum simulations are too numerous to list here and rapidly become anachronistic. Nevertheless, two outstanding examples include the simulation of Kibble-Zurek-type dynamics using a 69-qubit hybrid digital/analogue technique \cite{WOS:001427008900001} and a 5000-qubit quantum annealer \cite{WOS:000975071100001}.

Making effective use of quantum hardware now requires a clearer distinction between several operating regimes. Noisy physical-qubit devices can support proof-of-principle results, benchmarking studies, small-scale Hamiltonian simulation, and workflow development, but to date they have not by themselves established chemically useful advantage. Error mitigation can extend the reach of such devices by reducing bias in selected observables, but it generally introduces sampling and validation overheads that must be included in any assessment of usefulness.

A separate regime is emerging around logical qubits and early fault-tolerant computation. In this setting, quantum error correction is used to encode a more reliable logical qubit across many physical qubits, allowing deeper circuits than would be possible on uncorrected hardware. Early fault-tolerant devices are unlikely to resemble fully mature, large-scale fault-tolerant quantum computers. Their importance lies in whether they can support scientifically meaningful subroutines, such as more reliable phase-estimation or Hamiltonian-simulation primitives, at resource levels that remain far below the requirements for broad, general-purpose quantum computation. Concrete recent progress includes experimental realisations of logical operations with reduced error rates~\cite{googlequantumai2025colourcode}, and theoretical work that has substantially lowered the projected magic-state and overall algorithmic resource costs of fault-tolerant algorithms~\cite{gidney2024magicstate,gidney2025factor,zhao2026ultrahighrate}.

This distinction matters for molecular and materials modelling because the projected value of a quantum calculation depends strongly on the hardware regime. A noisy circuit, an error-mitigated utility experiment, an early logical-qubit calculation, and a fully fault-tolerant algorithm should not be discussed as if they support the same claims. For each proposed application, the relevant question is whether the available quantum resource can improve a defined scientific quantity, such as an energy difference, excitation energy, reaction barrier, spectrum, or descriptor, after state preparation, measurement, error treatment, and classical post-processing are included. In the quantum algorithms section that follows, we therefore distinguish between physical-qubit demonstrations, error-mitigated calculations, early fault-tolerant approaches, and fully fault-tolerant algorithms when discussing possible scientific value, and refer to this framework where relevant in subsequent sections.

\section{Quantum Algorithms}

A wide range of quantum algorithms has been proposed for applications in chemistry and materials science. A comprehensive review is beyond the scope of this article, but the algorithmic landscape has broadened substantially beyond the early focus on standard VQE and small-molecule examples. The relevant distinction is no longer simply between algorithms that are practical on NISQ hardware and those reserved for ideal fault-tolerant machines. It is more useful to ask what scientific role an algorithm can play, what hardware regime it requires, and what overheads must be included before a claim of advantage is meaningful.

Quantum phase estimation remains a central route to high-precision eigenvalue estimation for electronic-structure problems. In principle, QPE can provide systematically improvable energies when supplied with a state that has sufficient overlap with the desired eigenstate and when the molecular Hamiltonian can be implemented accurately. Its practical value, however, depends on deep circuits, coherent time evolution, state preparation, and error correction, placing it primarily in the early fault-tolerant or fully fault-tolerant regime rather than on present noisy hardware.

Variational methods, including VQE and adaptive variants such as ADAPT-VQE, remain important for near-term exploration because they use comparatively shallower circuits and allow part of the computational burden to be moved to a classical optimiser. Their limitations are now well understood: ansatz selection, optimisation instability, barren plateaus, measurement overhead, and sensitivity to noise can all dominate performance. For this reason, variational algorithms should be viewed as one family of hybrid methods rather than as the defining model for quantum chemistry on quantum computers.

A broader class of quantum-centric hybrid approaches has emerged between these extremes. Subspace and selected-configuration methods, including QSE, QSCI, SQD, quantum computed moments, and Krylov-type approaches, seek to extract useful spectral or energetic information from sampled quantum states while using classical post-processing to build effective low-dimensional representations. These methods are attractive because they can sometimes reduce circuit-depth requirements while retaining a connection to chemically meaningful quantities such as ground-state energies, excited states, transition properties, or response observables. Their practical value depends on whether sampling overheads, noise sensitivity, and classical reconstruction costs remain controlled for the target system. A related, generative strand instead trains a classical generative model to propose and refine candidate quantum circuits from measured energies, as in the generative quantum eigensolver (GQE)~\cite{nakaji2024generative}, offering an alternative to hand-designed ansätze for ground-state search.

For early fault-tolerant and fully fault-tolerant settings, modern Hamiltonian-simulation methods have reshaped resource estimates. Linear-combination-of-unitaries constructions, block encodings, qubitization, quantum signal processing, quantum singular value transformation, quantum walks, tensor-factorised Hamiltonians, and first-quantised algorithms provide more systematic routes to simulating molecular and materials Hamiltonians than early textbook constructions. These methods are especially relevant because they connect algorithmic scaling to explicit assumptions about basis size, Hamiltonian representation, precision, logical qubit counts, gate depth, and error-correction overheads.

The practical message is that no single algorithmic family defines the path to scientific value. NISQ and error-mitigated methods are useful for testing workflows, descriptors, and small active spaces; subspace and sampling-based methods may bridge part of the gap between noisy and logical-qubit devices; and block-encoding and phase-estimation-based methods provide the more viable route to high-precision quantum chemistry in the fault-tolerant regime. The choice of algorithm must therefore be tied to a target scientific observable and to a complete resource model, rather than to an abstract claim of quantum advantage.

\begin{table*}[t]
\centering
\caption{
Algorithm families relevant to molecular and materials modelling.
The table separates the likely hardware regime from the scientific role of each method.
The application-area column indicates where each family is most discussed in this article, not an exhaustive scope.
}
\label{tab:algorithm_map}
\renewcommand{\arraystretch}{1.15}
\begin{tabularx}{\textwidth}{p{0.18\textwidth} p{0.12\textwidth} p{0.13\textwidth} X X}
\toprule
\textbf{Algorithm family} &
\textbf{Typical regime} &
\textbf{Application area} &
\textbf{Scientific role} &
\textbf{Main unresolved issues} \\
\midrule

VQE and adaptive variational methods, including ADAPT-VQE &
NISQ / error-mitigated &
Chemistry, Materials Science, Biochemistry &
Active-space energy estimation, ansatz testing, and workflow prototyping on shallow circuits. &
Ansatz choice, optimisation stability, barren plateaus, measurement overhead, and noise sensitivity. \\

\addlinespace

Quantum subspace and sampling-based methods, including QSE, QSCI, SQD, quantum computed moments, and Krylov-type approaches &
NISQ to early FTQC &
Chemistry, Biochemistry &
Subspace reconstruction, excited-state estimation, spectral information, and improved energy estimates from sampled quantum states. &
Sampling cost, robustness to hardware noise, classical reconstruction cost, and scaling to larger active spaces. \\

\addlinespace

Quantum phase estimation and related eigenvalue-estimation methods &
Early FTQC / FTQC &
Chemistry, Materials Science, Biochemistry &
High-precision estimation of molecular energies and spectra, provided suitable input states and Hamiltonian simulation routines are available. &
State preparation, Hamiltonian implementation, logical-qubit requirements, circuit depth, and error-correction overhead. \\

\addlinespace

LCU, block encoding, qubitization, QSP/QSVT, and quantum-walk methods &
Early FTQC / FTQC &
Chemistry, Materials Science &
Systematic Hamiltonian simulation and modern resource estimation for chemistry and materials problems. &
Concrete circuit construction, ancilla overhead, precision targets, and error-correction cost. \\

\addlinespace

First-quantised and tensor-factorised algorithms &
FTQC &
Materials Science, Chemistry &
Improved asymptotic scaling for selected chemistry and materials problems, particularly where basis size and Hamiltonian representation dominate cost. &
Basis choice, data loading, state preparation, and implementation complexity. \\

\addlinespace

Quantum annealing and QAOA-style optimisation &
Analogue / NISQ / hybrid &
Materials Science &
Combinatorial search in crystal structure prediction, disorder, configurational analysis, and materials screening. &
Benchmarking against strong classical heuristics, hardware embedding overhead, constraint handling, connectivity, and repeated sampling. \\

\bottomrule
\end{tabularx}
\end{table*}

\section{Quantum computing challenges and perspectives}

Quantum computing is developing rapidly, with venture capital funding being poured into the sector, and it can be challenging to accurately evaluate the real progress being achieved in both quantum computing research and engineering. If we want to understand the true promise of quantum computing, we need to look at the actual experimental and theoretical advances.\\

Recent experiments on quantum processors show that controlled quantum systems can now perform tasks that are difficult to reproduce directly with classical simulation. Such experiments are scientifically important, but their relevance to chemistry and materials modelling depends on whether they can be connected to observables that matter for a real workflow. Random circuit sampling, analogue many-body dynamics, small electronic-structure calculations, and error-correction results should therefore be interpreted in different ways. They provide evidence of hardware and algorithmic progress, but they do not by themselves establish practical value for chemical prediction.\\

For molecular and materials modelling, useful quantum computing means improving a specified scientific decision relative to the best available classical workflow, after accounting for accuracy, uncertainty, runtime, energy cost, state preparation, measurement, error mitigation or correction, and classical post-processing. The current driver in classical HPC is to lower the energy cost per unit of compute \cite{di2024quantum,rallis2025interfacing} since the largest HPC facilities consume the electrical power of a small town, producing similar amounts of waste heat. Even with renewable electricity and innovative ways to repurpose the waste heat, we simply cannot expand just by building bigger \cite{nana2023energy}. One strategy is to switch to GPUs (graphics processing units) for the main processing power of the newest HPC facilities; GPUs consume less power for the same computation, but they are not suitable for every application, and are correspondingly harder to programme \cite{ahmadzadeh2024performance,elster2022nvidia}.  Thus, we are at a point where other forms of alternative hardware, such as quantum, are of serious interest, provided they are less power hungry. It is also important to note that classical computing has relied on hybrid architectures for a couple of decades now, handing off specialised tasks to dedicated processors \cite{mzukwa2024exploring}, which is where GPUs originated, and there are also separate chips on communications cards for internet connections and memory controllers, taking care of data compression and decompression, encryption, etc.  Other types of accelerators such as FPGAs (field programmable gate arrays) are also available to speed up common repetitive tasks.  It would be natural, therefore, to deploy quantum computers as quantum accelerators incorporated into the next generation of HPC \cite{silvano2025survey}. \\

There are multiple barriers to achieving useful quantum accelerator deployment \cite{venkatesha2024survey,mzukwa2024exploring}.  The first is that the basic theory of heterogeneous architectures is poorly developed.  We do not currently have a good understanding of how best to distribute the tasks between GPUs and CPUs to maximise the compute throughput and avoid processors idling while they wait for others to finish their calculations.  Second, the clock speeds of quantum hardware do not necessarily match those of classical hardware, and the mismatch can be several orders of magnitude.  Approaches to handle speed mismatch could include assigning many slow processors to work with one fast processor and parallelising the task for the slow processors, or dividing tasks at a higher level, so that the slow processor returns the result many fast clock cycles later when data exchange is needed \cite{song2024powerinfer,rosenfeld2022query}.  Third, the data types and encodings commonly used in quantum processors are not the same as in classical processors, where floating point numbers dominate.  Quantum processors usually work with integers, or amplitude encoding of numbers between zero and one.  Data conversion and quantum input state preparation are non-trivial tasks that must be included in the algorithm \cite{ilyas2024role}.  Likewise, reading out the result is not necessarily a simple measurement; the computation may require multiple repeats and post-processing of the sampled outputs. \\

Hardware development remains difficult, although incremental progress continues across several platforms \cite{daley2022practical}. Any computational advantage is expected to arise only for specific tasks, particularly those involving quantum dynamics, which are costly to simulate on classical computers. For this reason, quantum simulation of many-body systems is often viewed as the most likely early application, especially on specialised hardware; studies using quantum annealers, including work on phase transitions with D-Wave devices, illustrate this possibility \cite{miessen2025digital,fauseweh2024quantum}. More generally, progress depends on identifying parts of existing HPC algorithms where a quantum subroutine could be beneficial, rather than replacing entire workflows. Developing such targeted approaches before large-scale hardware is available remains valuable, provided the proposed quantum subroutine is benchmarked against strong classical baselines and evaluated as part of an end-to-end workflow.

\section{Early applications of quantum computing in molecular and materials modelling}

In his seminal piece, Feynman emphasised the inherent difficulty of simulating quantum systems with classical computers~\cite{Feynman82}. Computers able to exploit quantum effects were envisioned as a way of sidestepping this issue. Since then, the idea of using a controllable quantum system to study another potentially inaccessible system, also known as quantum simulation, has received a lot of attention~\cite{Georgescu14}. Specifically, considerable progress has been made in the computation of the ground state energy with a range of quantum computing architectures~\cite{Cao2019}. The current state of quantum simulation on existing devices ranges from small, highly controlled demonstrations to more ambitious studies of strongly correlated molecular fragments. Molecules such as Cr$_2$ remain valuable test cases because they stress classical electronic-structure methods through multireference character and strong correlation. They should, however, be treated as benchmarks for algorithmic and resource analysis rather than as evidence that practical quantum advantage has already been reached.

Claims about quantum advantage in such systems require precise qualification. A meaningful comparison must specify the active space, basis, target precision, electronic-structure quantity, classical baseline, quantum algorithm, encoding, circuit depth, measurement strategy, and whether the stated qubit count refers to physical or logical qubits. For example, a CASSCF active space such as (26,26) identifies the number of active electrons and orbitals, but it does not by itself define the resource requirement or the advantage threshold. The cost depends on the representation of the Hamiltonian, the target accuracy, the algorithmic route, and the error model.

For near-term devices, the more defensible role of such examples is to identify where quantum methods might eventually contribute: strongly correlated active spaces, excited states, reaction pathways, and quantum dynamical observables that remain challenging for classical methods. In the early fault-tolerant regime, the same examples can be used to test whether QPE-like, qubitization-based, or subspace methods provide a credible route to reducing uncertainty at acceptable resource cost. Until those assumptions are specified, phrases such as ``current threshold for quantum advantage'' should be avoided or replaced by narrower language such as ``a demanding benchmark for resource estimation in strongly correlated quantum chemistry.''\\

While much of the algorithmic development in quantum computing for molecules and materials has focused on improving the accuracy of energy estimation, early applications have demonstrated particular promise in the generation of candidate structures, which is especially evident in problems characterised by severe combinatorial complexity, such as crystal structure prediction (CSP) and disordered materials modelling, where quantum and quantum-inspired approaches have been investigated as ways to explore configurational spaces under constrained benchmark conditions~\cite{Gusev2023,AuYeung2023,Camino2023,Camino2025}. A unifying feature of these early applications is the reformulation of materials design problems as instances of combinatorial optimisation. In this approach, chemically meaningful degrees of freedom, such as atomic occupations, substitutions, or ordering patterns, are mapped onto binary decision variables, allowing the underlying materials problem to be expressed in an abstract mathematical form. When interatomic interactions are modelled using classical force fields or truncated cluster expansions, the resulting energy depends on pairs of such variables, leading naturally to a quadratic objective function.\\

Through standard transformations, constraints can be absorbed into the objective, yielding a Quadratic Unconstrained Binary Optimisation (QUBO) formulation. This mapping provides a clear conceptual bridge between chemical configuration space and optimisation problems that are amenable to both classical and quantum solution strategies~\cite{Yin2022,Gusev2023,Couzinie2024}. QUBO formulations are particularly relevant for quantum annealing platforms \cite{Kim2025}, such as those developed by D-Wave, which are designed to approximately solve Ising and QUBO problems by exploiting quantum dynamics~\cite{Berwald2018}. In this context, qubits represent binary configuration choices and their couplings encode effective interaction terms, enabling low-energy solutions to be interpreted as candidate material structures. Proof-of-principle demonstrations have shown that small-scale CSP and ordering problems can be addressed in this way~\cite{Gusev2023}, although current device limitations in qubit count, connectivity, and noise restrict accessible system sizes. More broadly, the same QUBO abstraction enables hybrid quantum–classical workflows and alternative quantum implementations, including gate-based approaches such as the QAOA, as well as specialised classical and analogue Ising machines~\cite{Ajagekar2020,Mohseni22,Abbas24}. While these methods do not yet guarantee access to true ground states for realistic materials, they offer a useful testbed for quantum-assisted configurational screening. Their practical value should be judged against strong classical heuristics, including the cost of embedding the problem on hardware, constraint handling, repeated sampling, and post-processing.

The possibility of quantum-accelerated biomolecular simulation, as originally proposed by Kendon and Harris~\cite{Kendon2010}, remains a long-term aspiration. A breakthrough in quantum computing for molecular quantum chemistry applications would immediately have strong implications for biomolecular modelling; however, as better quantum simulations would enable us to model enzyme catalysis, which is fundamental to biology at the molecular level~\cite{Baiardi2023}. Furthermore, the potential to use quantum computing for applications related to the optimisation of networks might be helpful in systems biology, which considers how the many proteins interacting in metabolic networks give rise to complex behaviour~\cite{Kitano2002,Cordier2022}. Quantum computers are noisy, but so is biology. With a careful choice of algorithm, this noise may potentially be harnessed as part of the simulation, rather than treated solely as an error source~\cite{Dorner2012}. If noisy quantum simulations are fast enough, then their inaccuracy would not be an insurmountable problem, so long as the correct solution is the most probable~\cite{AuYeung2023}. In the meantime, hybrid approaches, in which quantum computing is used selectively for the most challenging components of a simulation and coupled to classical computation for the remainder, may therefore offer a practical route towards impact in biomolecular modelling and related design problems.\\

\section{Conclusions}
Quantum computing brings back a basic question in chemical modelling: what becomes possible if we can treat electronic correlation and quantum dynamics with fewer approximations than on classical hardware? The potential is genuine, but it will matter only for a subset of problems - those where quantum mechanics is the limiting factor in practice, not just in principle. \\

Quantum computers are unlikely to displace the methods that dominate chemical and materials modelling. A more realistic outlook is that they will be used as specialised components within established workflows. In chemistry, the clearest route is to improve the treatment of correlated electronic structure in carefully chosen regions of a system, often within embedding schemes alongside the use of higher-quality reference calculations to assess and improve approximate methods such as density functionals, force fields, and machine-learned potentials. In materials science, early value may come from improving search over structure and composition when the main difficulty is exploring a large space of possibilities rather than performing a single high-accuracy energy calculation. In biochemistry, near-term use is likely to be local: treating reactive or strongly polarised regions more accurately inside multiscale models, while wider claims about accelerating biomolecular simulation must be tested end-to-end.\\

These considerations shift what the field should prioritise. Progress should be judged using benchmarks that reflect real use: error bars on energies and free energies that translate into errors in rates, selectivities, phase stability, or spectra; validation against experiments and high-level classical reference calculations; and demonstrations that include the full costs of state preparation, measurement, and coupling to classical simulation. The practical challenge is not only building larger devices but also making them usable within heterogeneous computing environments, where data formats, runtimes, and error behaviours differ sharply from those in classical systems.\\

The next stage of progress should be judged less by isolated proof-of-concept experiments and more by whether quantum calculations reduce uncertainty in scientific workflows that already matter. For chemistry and materials science, that means clearer benchmarks for correlated active spaces, excitation energies, reaction barriers, rates, spectra, and configurational search. It also means reporting the full computational cost, including state preparation, measurement, error handling, classical post-processing, and integration with existing HPC and multiscale methods.


\section{Acknowledgements}

TWK would like to thank Vendel Szeremi, Mateusz Meller, Alberto Popescu and Joseph Thacker at STFC and our collaborators at the National Quantum Computing Centre, Riverlane and SEEQC, and acknowledges funding from the InnovateUK project “Quantum Enhanced Computing Platform for Pharmaceutical R\&D – QuPharma” (project number 10005792). Support was also provided by the EPSRC Hub for Quantum Computing via Integrated and Interconnected Implementations (QCi3, grant EP/Z53318X/1), with funding attributed to PAW, VK, and SH. Further support was provided by the CCP-QC programme (grant EP/T026715/2) and by the UKRI Digital Research Infrastructure (DRI) and STFC-funded CCP-QC Bridge Project, \textit{Extended Case Studies for Neutral Atom Hardware}. CCP-QC funding is attributed to BC, CRAC, SH, SMW, JM, and AME. Additional support was received from the QEVEC programme (EP/W00772X/2), with funding attributed to BC, SW, JB, VK, and CRAC. In addition, BC, AME, PAW, JAN, TWK, CRAC and SMW are supported by UKRI EPSRC grant numbers EP/W026775, EP/X035859, EP/Z53318X, and UKRI2710.

\newpage
\bibliographystyle{rsos}
\bibliography{references}

\end{document}